\documentclass[journal]{IEEEtran}
\usepackage{color}
\usepackage{xcolor,colortbl}
\usepackage{graphicx}
\usepackage{tikz}
\usepackage{tikzscale}
\usetikzlibrary{arrows.meta}
\usepackage{pgfplots}
\usepackage{dashrule}
\usepackage{caption}
\usepackage{subcaption}
\usepackage{multicol}
\usepackage{balance}
\usepackage{diagbox}
\usepackage{multirow}
\usepackage{hanging}
\usepackage{ragged2e}
\usepackage{romannum}

\usepackage{listings}

\usepackage{url}
\usepackage{xurl}
\usepackage{hyperref}
\usepackage{cite}

\usepackage{algorithmic}
\usepackage[titlenumbered,ruled]{algorithm2e}

\usepackage{soul}       
\usepackage{textcomp}   
\usepackage{comment}

\usepackage{amsmath,amssymb,amsfonts}

\usepackage{accents}     

\newcounter{substep}

\ifCLASSINFOpdf
\else
\fi
\begin{document}
%
\title{Low-Power On-Device Gesture Recognition with Einsum Networks}
%
%
%
%

\author{Sahar~Golipoor,
        Lingyun~Yao, Martin~Andraud, and~Stephan~Sigg
\thanks{We acknowledge funding by the European Union in the frame of the Horizon Europe EIC project SUSTAIN (project no. 101071179), and Holden (project no. 101099491). Views and opinions expressed are those of the authors and do not necessarily reflect those of the European Union.}
\thanks{Sahar~Golipoor and Stephan~Sigg are with the Department of Information and Communications Engineering, Aalto University, Espoo, 02150 Finland  (e-mail: \{sahar.golipoor, stephan.sigg\}@aalto.fi).}

\thanks{Lingyun~Yao is with the Department of Electronics and Nanoengineering, Aalto University, 02150 Espoo, Finland (e-mail: lingyun.yao@aalto.fi).}

\thanks{Martin~Andraud is with UCLouvain, ICTEAM, Louvain-La-Neuve, Belgium, and also with the Department of Electronics and Nanoengineering, Aalto University, 02150 Espoo, Finland (e-mail: martin.andraud@uclouvain.be).}
}

\maketitle
\sethlcolor{blue!50!white}

\begin{abstract}
We design a gesture-recognition pipeline for networks of distributed, resource constrained devices utilising Einsum Networks. 
Einsum Networks are probabilistic circuits that feature a tractable inference, explainability, and energy efficiency.
The system is validated in a scenario of low-power, body-worn, passive Radio Frequency Identification-based gesture recognition. 
Each constrained device includes task-specific processing units responsible for Received Signal Strength (RSS) and phase processing or Angle of Arrival (AoA) estimation, along with feature extraction, as well as dedicated Einsum hardware that processes the extracted features. 
The output of all constrained devices is then fused in a decision aggregation module to predict gestures. 
Experimental results demonstrate that the  method outperforms the benchmark models.
\end{abstract}





%
\IEEEpeerreviewmaketitle

\section{Introduction}\label{sec:introduction}
\IEEEPARstart{T}{he} power consumption of recognition pipelines places a significant constraint on the domains in which they may be employed~\cite{kellogg2014bringing}.
For instance, smart glasses, watches, smartphones~\cite{bhattacharyya2024helios,varposhti2024energy,vostrikov2024unsupervised}, but also, for example, smart home systems~\cite{catanua2024smart,thirunavukkarasu2025enhanced}, where power consumption is a significant concern (Buildings consume close to 50\% of total primary energy~\cite{cao2016building}).  

Ultra-low-power processing units may help to realize recognition in such power-constrained domains, however, these processors have severe constraints, which limits their capability to support best performing recognition pipelines~\cite{benatti2019online,zhang2020ultra,scrugli2024semg}. 

We design and optimize a gesture recognition pipeline for a highly optimized acceleration module, utilizing probabilistic reasoning, namely Einsum Network operations. 

Particularly, to achieve a minimum overall energy budget, we envision an environmental recognition system in which body-worn ambient backscatter tags are read out and their stimuli are interpreted through the probabilistic reasoning pipeline that is executed on the acceleration module. 

Since ambient backscatter operates without a dedicated carrier emitter, the overall power consumption of the system is minimal~\cite{duan2020ambient,wu2022survey}. 
For instance recent work explored reader systems with on-board processing~\cite{hu2023fully,liao2025enable}.

In general, radio-based sensing has been employed for localization, motion and emotion 
recognition~\cite{shahbazian2023human,miao2025wi,kong2024survey,zahid2024comprehensive}, for instance, for assisted living, healthcare, and human-robot interaction~\cite{wen2024survey}. 
Key benefits of radio sensing are its robustness against both poor lighting and occlusion. 
Despite the distinct privacy issues related to its ability to sense through obstacles and capture subtle movements, RF sensing systems do not record visual data~\cite{windl2025privacy}.

Recently proposed recognition models for RF sensing are constrained by their computational complexity~\cite{salami2022tesla,palipana2021pantomime}, which worsens both energy consumption and real-time responsiveness~\cite{chen2019deep}. 

We propose to combine gesture recognition via Einsum Networks~\cite{peharz2020einsum} on distributed, resource constrained devices with ambient backscattering as a physical sensing modality~\cite{yin2020back}, as it allows highly power-efficient environmental sensing~\cite{zargari2023signal}. 
Einsum Networks, since they can be parallelized in hardware, are a class of computationally efficient and low-power Probabilistic Circuits which guarantee tractable inference.


Our contributions are as follows:
\begin{itemize}
\item We present a low-power gesture recognition model that achieves 97.96\% accuracy, surpassing all Deep Neural Network (DNN) baselines while reducing computation by at least 15\%. Compared to the random forest classifier, our model reduces computation by 8.34\% and is hardware-efficient.
\item We propose a structure where constrained devices independently process data and train their own Einsum Networks with diverse feature extraction strategies as weak learners, which are then combined to boost overall performance.
\item We demonstrate gesture recognition using body-worn tags, leveraging Received Signal Strength (RSS), phase, and Angle of Arrival (AoA) measurements.
\end{itemize}

\section{Related Work}\label{related work}
Ambient backscattering allows the design of low-power sensing systems since ambient radiation is utilized passively for the communication and sensing system~\cite{kellogg2014bringing}. 
Specifically, the use of backscatter-type tags for the recognition of human-object interaction, activity or gesture recognition has been studied prominently~\cite{li2015idsense}, as well as customer behavior~\cite{zhou2017design}, detection of emotional responses~\cite{shangguan2017enabling}, and observation of collaborative actions~\cite{li2016deep}. 
The authors in~\cite{bu2018rf}, for instance, turned physical items into interactive human-computer interfaces by attaching tag arrays to objects and using orthogonal antennas for motion tracking. 

Contrary to the above work that utilize stationary backscatter tags integrated into environments and objects, we propose to integrate them into garment so that tags can be interpreted as markers attached to various body parts as in~\cite{salo2024carbon, kruse2024smart}. 

These markers enable the differentiation of movement from distinct body parts~\cite{golipoor2023accurate}.
In~\cite{golipoor2024rfid}, we introduced a wearable tag-based system capable of recognizing activities by jointly analyzing features derived from both phase and RSS and other groups have demonstrated gesture recognition through backscatter tags embedded in gloves~\cite{cheng2019air, xie2017multi} as well as position-invariant sign language recognition ~\cite{zhang2023rf} by normalizing the horizontal rotation angle and radial distance.

Movement-induced tag rotations cause polarization mismatches and degraded signal quality~\cite{clarke2006radio}. 
Additionally, anatomical features and tissue composition affect signal reflections~\cite{vasisht2018body} cause variations that alter gesture signatures and reduce recognition accuracy~\cite{yu2019rfid}. 

In our previous work~\cite{golipoor2024environment}, we developed two models, processing IQ scatter plots in a VGG16 architecture as well as characteristic phase patterns. 
In this present study, we incorporate AoA tracking as new feature to further improve the gesture recognition models. 

Deep learning approaches have achieved impressive gesture recognition accuracy, however, their growing model size as well as limited explainability are increasingly concerning~\cite{marcus2020next}. 

We explore a probabilistic approach to gesture recognition. 
Probabilistic models are explainable, due to explicit probability calculations, in contrary to black-box neural networks~\cite{GZ15} and particularly Probabilistic Circuits feature tractable inference~\cite{Ventola2023}.
This is crucial to human data-related AI applications. 

\begin{figure}
\centering
\includegraphics[width=0.85\linewidth]{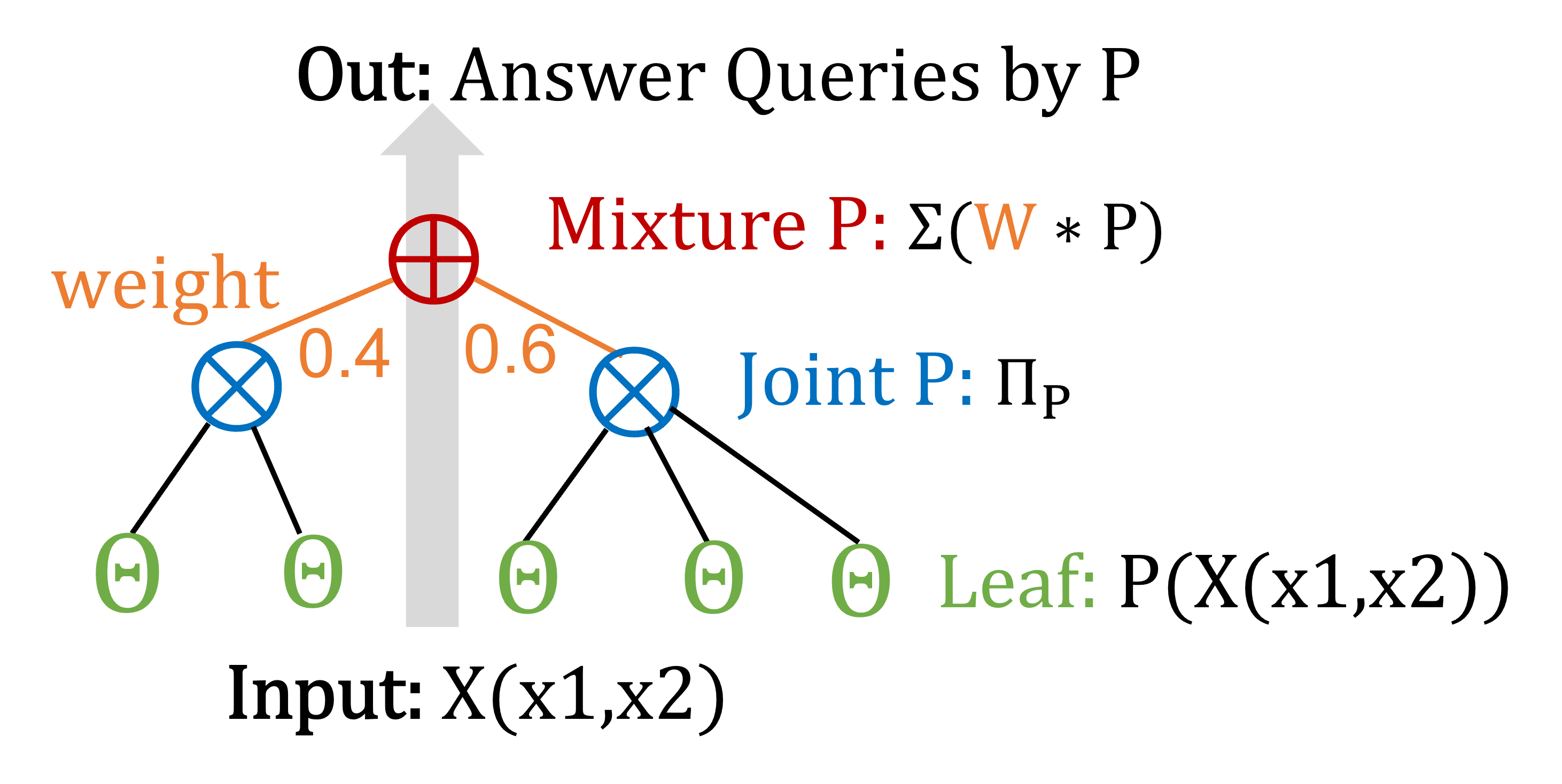}
\caption{Example structure of Probabilistic model: Probabilistic circuits; Leaf constrained devices $\Theta$ encodes distributions; $\otimes$ and $\oplus$ perform joint and mixture probability respectively.}
\label{PCs}
\end{figure}

Fig.~\ref{PCs} shows a Probabilistic Circuit for $\mathbf{X}=(x_1,x_2)$. 
The leaf constrained devices $\Theta$ store distributions for $\mathbf{X}$. 
A product node $\otimes$ computes joint probability by multiplying its disjoint scopes: 
$p(\mathbf{x})= \prod_k p_k(\mathbf{x})$. 
A sum node $\oplus$ produces a mixture distribution: $p(\mathbf{x})=\sum_k w_k\,p_k(\mathbf{x})$ with $\sum_k w_k=1$. 
The output of queries represent probabilities computed via bottom-up propagation in the learned probabilistic circuit, without the need for a normalization layer. 
Moreover, due to structural constraints, Probabilistic Circuits enable polynomial complexity inference, referred to as tractable inference, which provides an advantage for resource-constrained and edge applications~\cite{choi2020probabilistic}.

A recently proposed probabilistic circuit model, Einsum Networks, adopts a neural network style structural representation while preserving the core probabilistic properties of Probabilistic Circuits~\cite{peharz2020einsum}.
This regular and grid-like architecture enables more efficient computation compared to earlier, irregular probabilistic models.

\section{System and Signal Models}
We introduce the system model, which includes the reader, constrained devices, and the decision aggregator module, along with the received signal model that will be used in the gesture recognition task.

\subsection{System Model}
The reader consists of a two-element antenna array positioned at a known location, denoted by $\mathbf{p}{_\text{R}} \in \mathbb{R}^2$. The environment also includes one or more passive single-antenna tags,\footnote{Passive tags are powered by the signal transmitted from the reader.} whose positions are unknown. These positions are denoted as $\mathbf{p}{_\text{i}} \in \mathbb{R}^2 $, where $i \in {1, \dots, N_t}$, and $N_t$ denotes the number of tags. The tags must be within the field of view of the antennas to avoid misdetection.\footnote{Tag misdetection also depends on the transmit power, which can be configured through the reader's software.} The reader transmits a carrier wave (CW) using right-hand circular polarization to prevent misdetection caused by polarization mismatch between the reader's antennas and the tags' antennas. Additionally, we assume that the system is configured such that the tags operate within the far-field range of the reader's antenna, where $\| \mathbf{p}{_\text{R}}-\mathbf{p}{_\text{i}} \| \ge 2D^2/\lambda$ with $D$ being the maximum dimension of the reader's antenna array~\cite{sherman1962properties}.


We utilize Alien AZ 9662 passive tags, attaching two tags to the back of each hand. 
The setup employs an Impinj Speedway R420 reader~\cite{impinj_r420} and two Vulcan PAR90209H antennas with $9$ dBic antenna gain as well as elevation and azimuth beamwidth of $70^{\circ}$. 
The antennas are connected to a laptop running the ItemTest software for communication between the reader and the system.

The global coordinate system is aligned with the reader’s, which employs a two-element uniform linear array with elements at $(-d/2,0)$ and $(d/2,0)$. The spacing between the elements is assumed to be $d = 0.8\lambda$, where $\lambda$ is the wavelength of the carrier's frequency. The reader utilizes Smart Antenna Switching~\cite{aiouaz2012rfid}. 

We deploy multiple constrained devices, each consisting of a processor and a Probabilistic Circuits graph. 
These constrained devices process the signals, extract features, and perform classification using probabilistic reasoning. 
The final prediction is achieved by averaging the class scores from all constrained devices.
In our instrumentation to demonstrate the feasibility on distributed constrained devices, the distinct features processed at distributed constrained devices are collected from a single reader. 
In an actual implementation, each constrained device reads data from an independent reader, processes it locally, and performs on-board classification.

\subsection{Signal Model}
The passive tag consists of an integrated circuit (IC) and an antenna, which is made up of two distinct segments forming a dipole. During reader-tag communication, the reader transmits a CW signal, and the tag's IC harvests power from the received signal. It then modulates the response by changing the antenna's load to backscatter or absorb the wave from the reader, resulting in the reflection coefficient, $\Gamma \in \{0,1\}$.\footnote{The reflection coefficient is calculated as : $\Gamma = \frac{Z_\text{Ant}-Z_\text{Load}}{Z_\text{Ant}+Z_\text{Load}}$, where $Z_\text{Ant}$ and $Z_\text{Load}$ denote the antenna impedance (typically around $50\Omega$) and the load impedance controlled by the tag's IC, respectively. When $Z_\text{Load}$ is matched with $Z_\text{Ant}$, i.e., $Z_\text{Load}= Z_\text{Ant}$, then  $\Gamma=0$, meaning that there is no reflection. If the $Z_\text{Load}=0$, i.e., a shortcut circuit, then $\Gamma=1$. In practice, due to imperfect load matching, the maximum value of the $\Gamma$ is close to $1$ and the minimum value is close to zero.}The system is a smart antenna switching system and samples the backscattered signal from the tags at each antenna element at different times, but periodically. 
The reader outputs the backscattered baseband signal, i.e., the in-phase and quadrature components (IQ).


 The AoA in the azimuth direction of the line-of-sight (LoS) path from the $i$-th tag relative to the center of the reader's antenna array is calculated as $\theta_i =   \text{atan2}\left( [\mathbf{q}_i]_2, [\mathbf{q}_i]_1 \right)- \pi/2$, where $\mathbf{q}_i = \frac{\mathbf{p}_i - \mathbf{p}_\text{R}}{\| \mathbf{p}_i - \mathbf{p}_\text{R} \|}$. 
 Similarly, the AoA of the $\ell$-th non-line-of-sight (NLoS) path associated with the $i$-th tag, which originates from the $\ell$-th scatterer located at $\mathbf{p}_{s_\ell}$, is computed as $\theta_i^\ell = \text{atan2}\left( [\mathbf{q}_i^\ell]_2, [\mathbf{q}_i^\ell]_1 \right)-\pi/2 $, where $\mathbf{q}_i^\ell = \frac{\mathbf{p}_{s_\ell} - \mathbf{p}_\text{R}}{\| \mathbf{p}_{s_\ell} - \mathbf{p}_\text{R} \|}$.
Therefore, under the above assumptions, the received signal at the $m$-th antenna of the reader, backscattered from the $i$-th tag, is denoted by $y_\text{m,i} (k) \in \mathbb{C}$ as:
\begin{align}\label{eq:ym}
    y_\text{m,i} (k) &= g_i \sqrt{P} \, a_\text{m}(\theta_\text{i}) \, s_\text{m}(k) 
    + \sum_{\ell = 1}^{L} g^{\ell}_i \sqrt{P} \, a_\text{m}(\theta^{\ell}_\text{i}) \, s_\text{m}(k) \nonumber \\ 
    &+ \nu_\text{m,i}(k), 
    \quad \forall m,i \in \{1, 2\}, \; \forall k \in \{1, 2, \dots, K\}
\end{align}
where $g_i$ and $g^{\ell}_i$ represent the channel gains of the round-trip path between the reader and the $i$-th tag for the LoS and the $\ell$-th NLoS paths, respectively. 
$s_\text{m}(k)= x(4k+ 2m+i-6)$ is the transmit signal over $m$-th antenna at time $k$, and $x(k)=e^{\jmath 2\pi f_\text{c} k T_\text{s} }$, $f_\text{c}$ is the the carrier frequency of the transmitted signal from the reader, $T_\text{s}$ is the sampling period of the ADC.  $\nu_\text{m,i}(k)$ indicates the effect of additive thermal noise at the reader,
 modeled as a zero-mean Gaussian distribution with variance $\sigma^2$. $P$ is the transmit power at the reader in the transmission mode. $L$ denotes the total number of NLoS paths. The $m$-th element of the steering vector is expressed as $a_\text{m}(\theta_\text{i})= \text{exp}(\jmath \frac{4\pi d}{\lambda } (m-1) \sin(\theta_\text{i}) ) \quad \forall m\in \{1,2\}$.\footnote{Unlike the standard steering vector mentioned in the literature, which is $a_\text{m}(\theta_\text{i})= \text{exp}(\jmath \frac{2\pi d}{\lambda } (m-1) \sin(\theta_\text{i}) ) \quad \forall m\in \{1,2\}$, the phase difference in the considered system is twice that of the one mentioned in the literature. This is due to the round-trip path between the reader and the tag.}
 We assume that the NLoS paths are weaker than the LoS path, which allows us to focus on the tag's AoA estimation in the subsequent step. Thus, we rewrite~\eqref{eq:ym} as
\begin{align}\label{eq:ym_modified}
    y_\text{m,i} (k) &= g_i \sqrt{P} \, a_\text{m}(\theta_\text{i}) \, s_\text{m}(k) + \nu_\text{m,i}^{\prime}(k), \\ 
    &\quad \forall m,i \in \{1, 2\}, \; \forall k \in \{1, 2, \dots, K\}, \nonumber
\end{align}
where
$\nu_\text{m,i}^{\prime}(k) \triangleq \sum_{\ell = 1}^{L} g^{\ell}_i \sqrt{P} \, a_\text{m}(\theta^{\ell}_\text{i}) \, s_\text{m}(k) + \nu_\text{m,i}(k)$
represents the aggregated interference from NLoS paths and additive noise.

For the next step, we concatenate all the observations from each antenna element, as given in (\ref{eq:ym_modified}), over the observation time, i.e., $k\in \{1, \dots, K\}$. Thus, we have:
 
\begin{align}\label{eq:ym-vec}
    \mathbf{Y}_i = g_i \sqrt{P}  \boldsymbol{A}(\theta_\text{i}) \mathbf{S} + \boldsymbol{N}_i\quad \forall i \in \{1, 2\},  
\end{align}
where $\mathbf{Y}_i= \begin{bmatrix} \mathbf{y}_\text{1,i}^\top \\ \mathbf{y}_\text{2,i}^\top \end{bmatrix} \in \mathbb{C}^{2 \times K/4}$, $\boldsymbol{A}(\theta_\text{i})=\text{diag}(\boldsymbol{a}(\theta_\text{i}))  \in \mathbb{C}^{2\times2}$, $\boldsymbol{a}(\theta_\text{i})= [1\quad \text{exp}(\jmath \frac{4\pi d}{\lambda } (m-1) \sin(\theta_\text{i}) ) ]^\top $ $\mathbf{S} = \begin{bmatrix} \mathbf{s}_\text{1,i}^\top \\ \mathbf{s}_\text{2,i}^\top \end{bmatrix} \in \mathbb{C}^{2 \times K/4}$, where $\mathbf{s}_\text{1,i}$ and $\mathbf{s}_\text{2,i}$ are the transmit signals over the first and second antennas to the $i$-th tag, respectively. Also, $\mathbf{y}_\text{1,i} \in \mathbb{C}^{K/4}$ and $\mathbf{y}_\text{2,i} \in \mathbb{C}^{K/4}$ represent all the IQ samples from the first and second antenna elements from the $i$-th tag, respectively, over the observation time. Similarly, $\boldsymbol{N}_i = \begin{bmatrix} \boldsymbol{\nu^{\prime}}_\text{1,i}^\top \\ \boldsymbol{\nu^{\prime}}_\text{2,i}^\top \end{bmatrix} \in \mathbb{C}^{2 \times K/4}$, where $\boldsymbol{\nu^{\prime}}_\text{1,i} \in \mathbb{C}^{K/4}$ and $\boldsymbol{\nu^{\prime}}_\text{2,i} \in \mathbb{C}^{K/4}$ denote all the interference and noise samples from the first and second antenna elements during the reception from the $i$-th tag, respectively, over the observation time. 

\section{Data Collection and Processing}
\begin{figure*}
\begin{subfigure}{0.125\textwidth}
    \includegraphics[width=\linewidth]{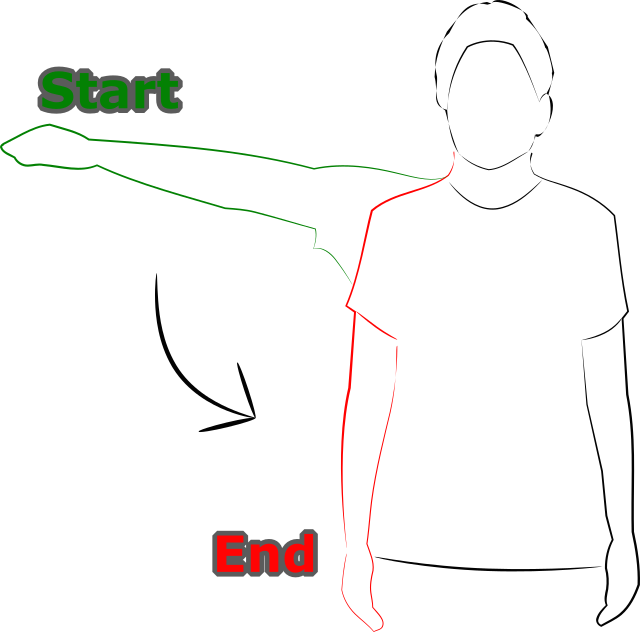}
    \caption{}
    \label{g}
\end{subfigure}
\begin{subfigure}{0.12\textwidth}
    \includegraphics[width=\linewidth]{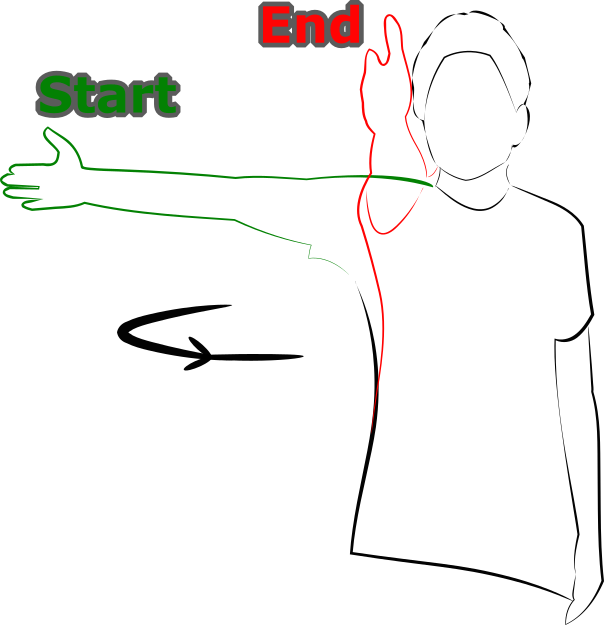}
    \caption{}
    \label{g}
\end{subfigure}
\begin{subfigure}{0.125\textwidth}
    \includegraphics[width=\linewidth]{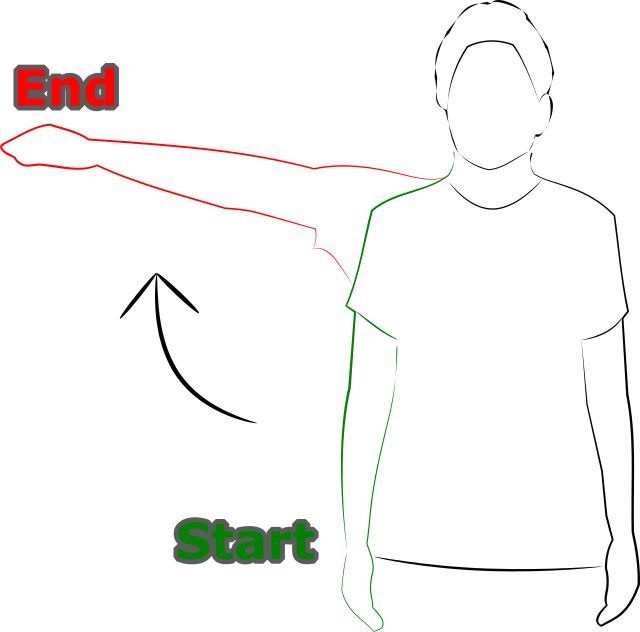}
    \caption{}
    \label{g}
\end{subfigure}
\begin{subfigure}{0.09\textwidth}
    \includegraphics[width=\linewidth]{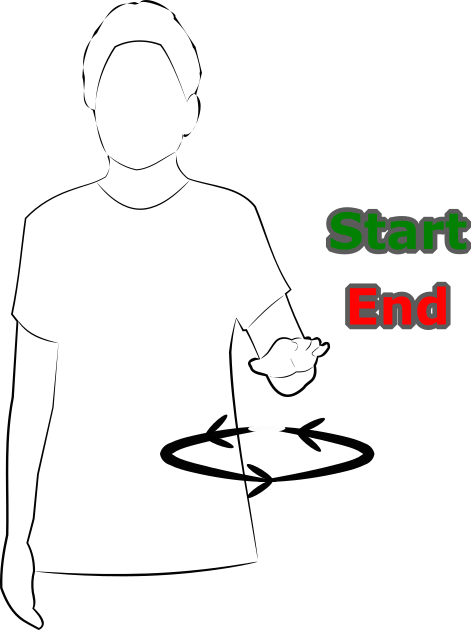}
    \caption{}
    \label{v}
\end{subfigure}
\begin{subfigure}{0.095\textwidth}
    \includegraphics[width=\linewidth]{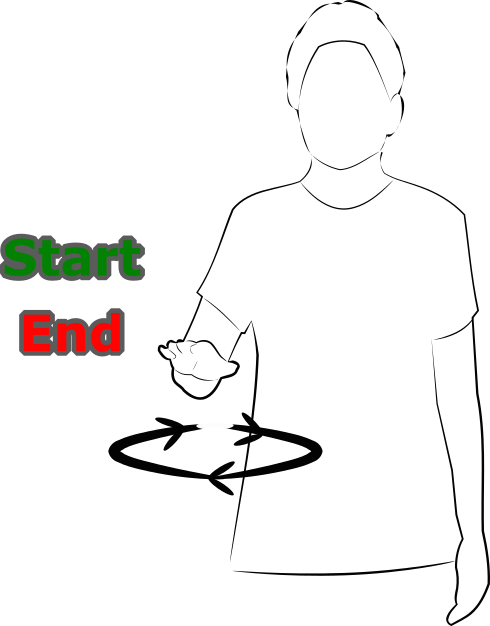}
    \caption{}
    \label{v}
\end{subfigure}
\begin{subfigure}{0.085\textwidth}
    \includegraphics[width=\linewidth]{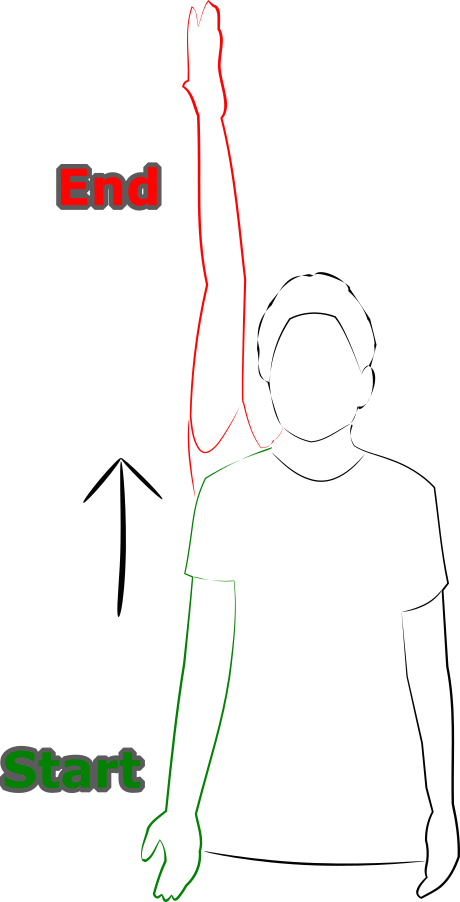}
    \caption{}
    \label{v}
\end{subfigure}
\begin{subfigure}{0.085\textwidth}
    \includegraphics[width=\linewidth]{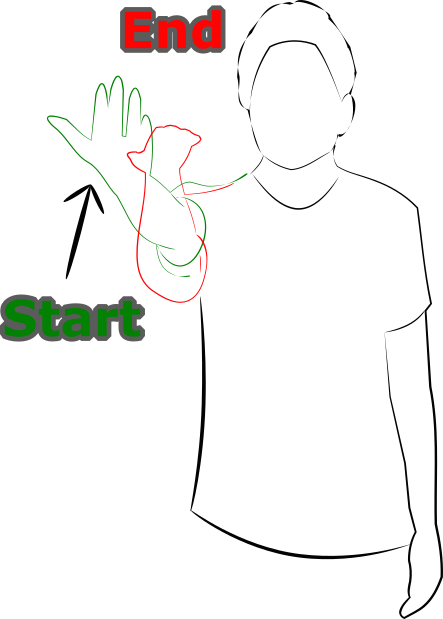}
    \caption{}
    \label{v}
\end{subfigure}
\begin{subfigure}{0.095\textwidth}
    \includegraphics[width=\linewidth]{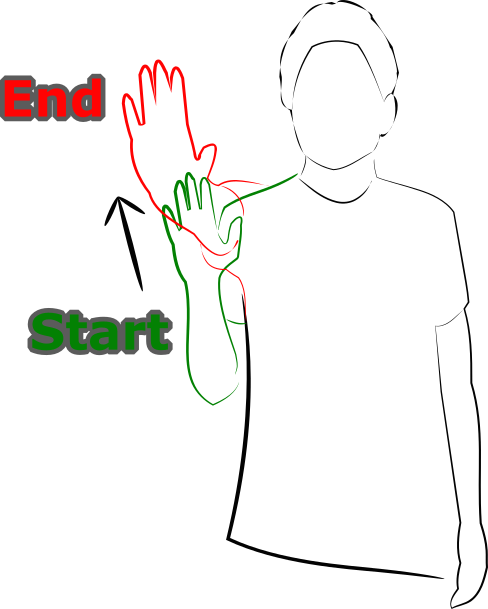}
    \caption{}
    \label{v}
\end{subfigure}
\begin{subfigure}{0.145\textwidth}
    \includegraphics[width=\linewidth]{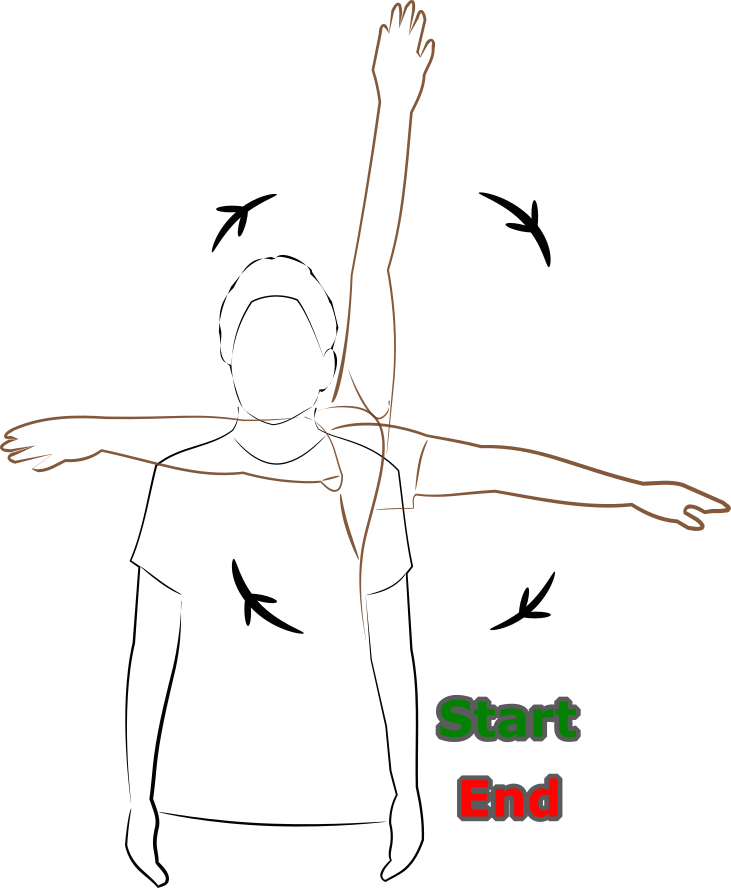}
    \caption{}
    \label{v}
\end{subfigure}
\begin{subfigure}{0.145\textwidth}
    \includegraphics[width=\linewidth]{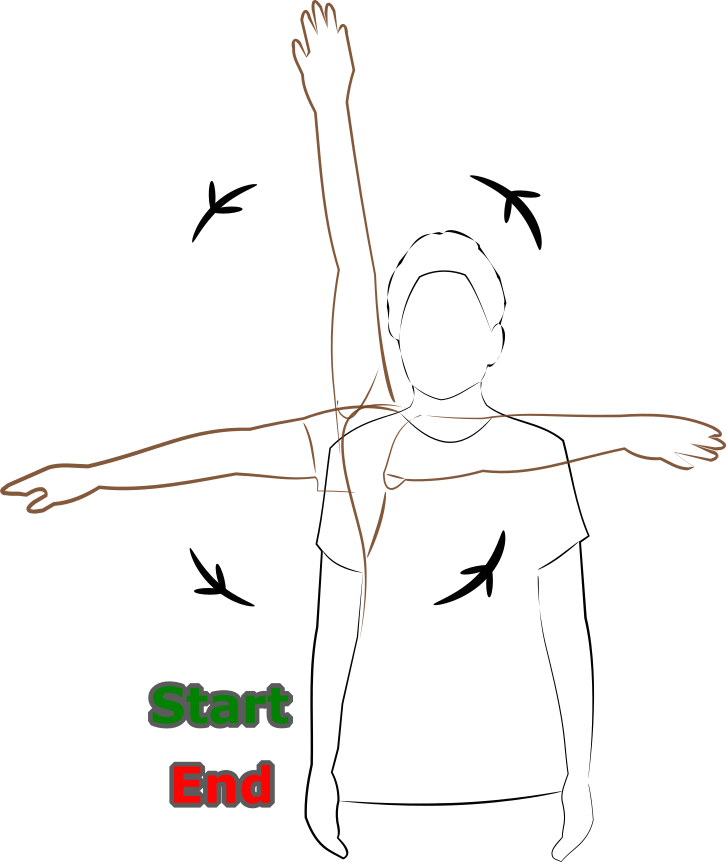}
    \caption{}
    \label{v}
\end{subfigure}
\begin{subfigure}{0.133\textwidth}
    \includegraphics[width=\linewidth]{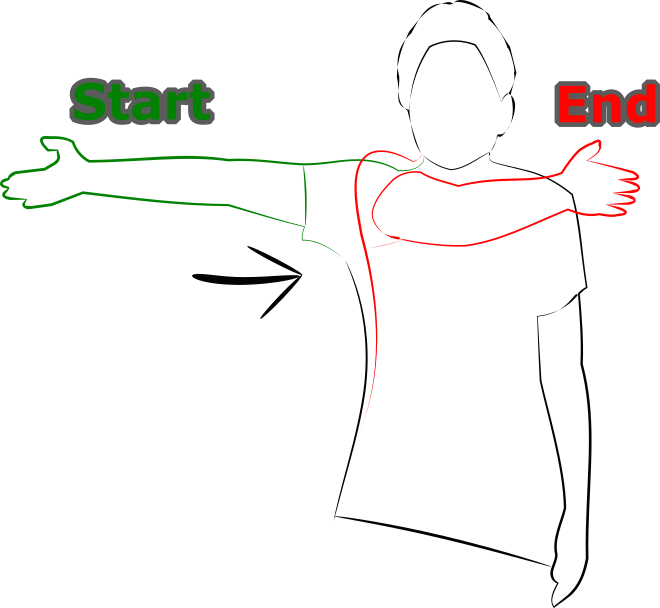}
    \caption{}
    \label{v}
\end{subfigure}
\begin{subfigure}{0.138\textwidth}
    \includegraphics[width=\linewidth]{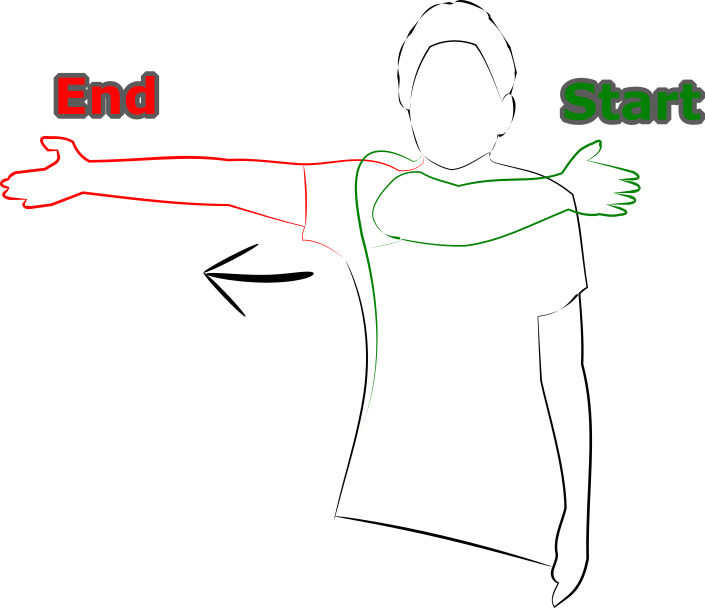}
    \caption{}
    \label{v}
\end{subfigure}
\begin{subfigure}{0.183\textwidth}
    \includegraphics[width=\linewidth]{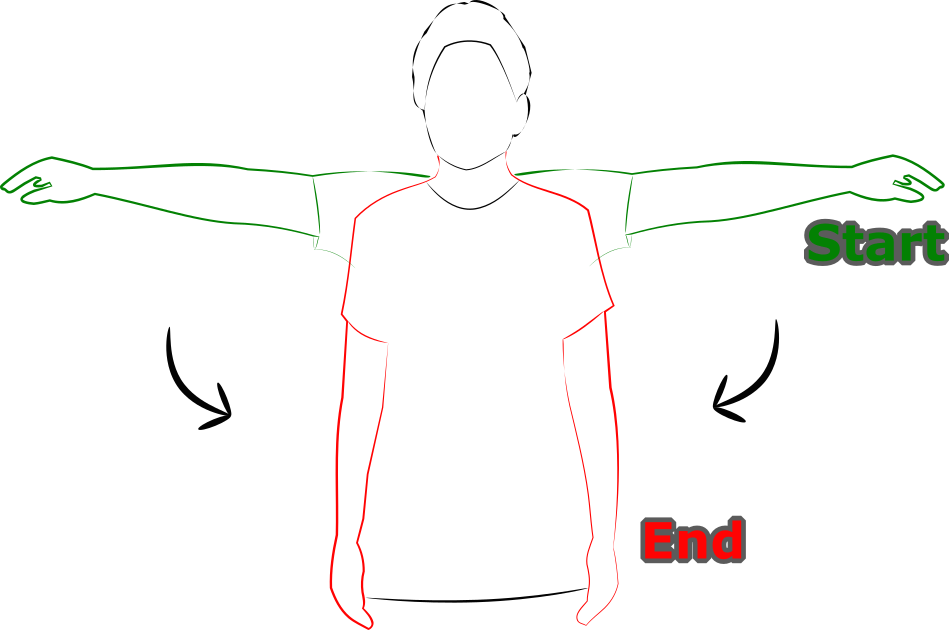}
    \caption{}
    \label{v}
\end{subfigure}
\begin{subfigure}{0.188\textwidth}
    \includegraphics[width=\linewidth]{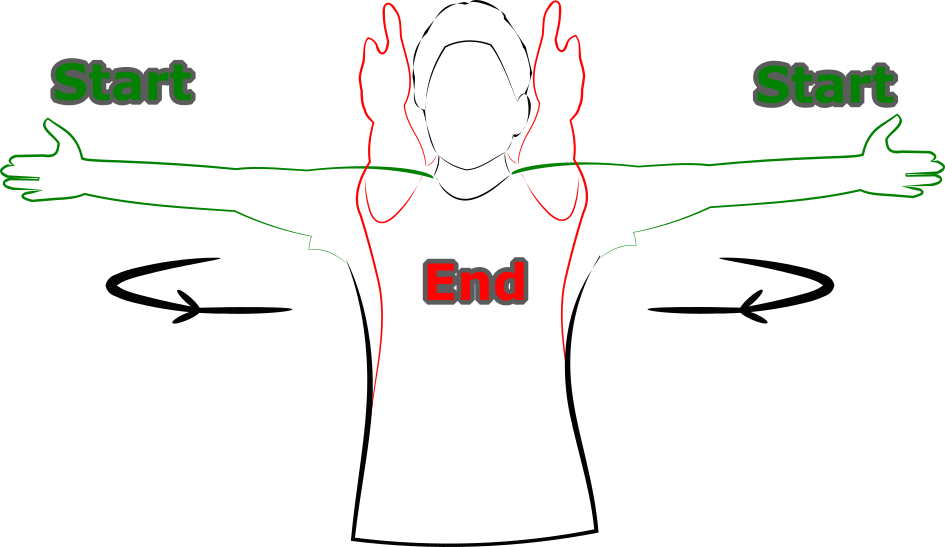}
    \caption{}
    \label{v}
\end{subfigure}
\begin{subfigure}{0.19\textwidth}
    \includegraphics[width=\linewidth]{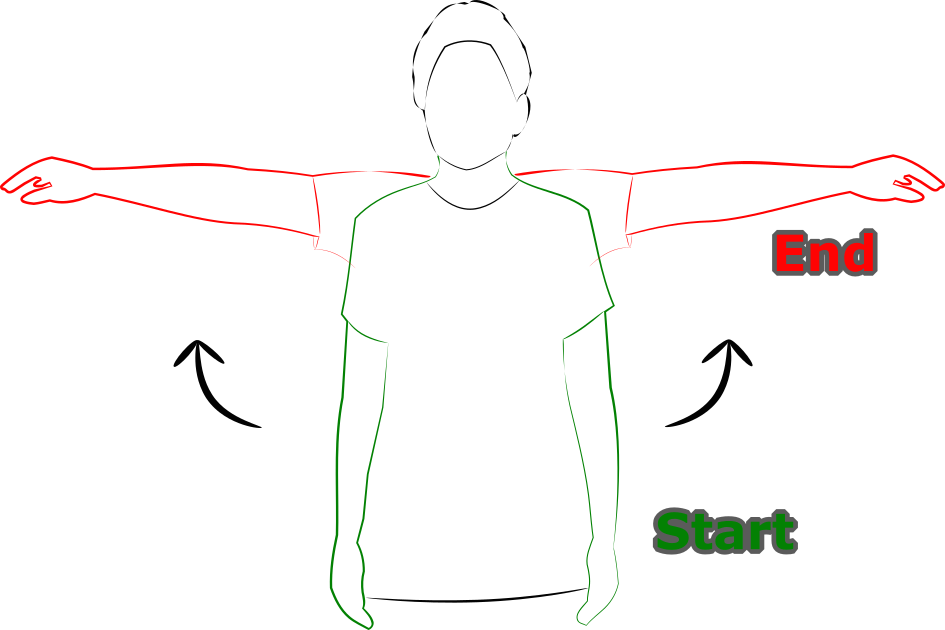}
    \caption{}
    \label{v}
\end{subfigure}
\begin{subfigure}{0.14\textwidth}
    \includegraphics[width=\linewidth]{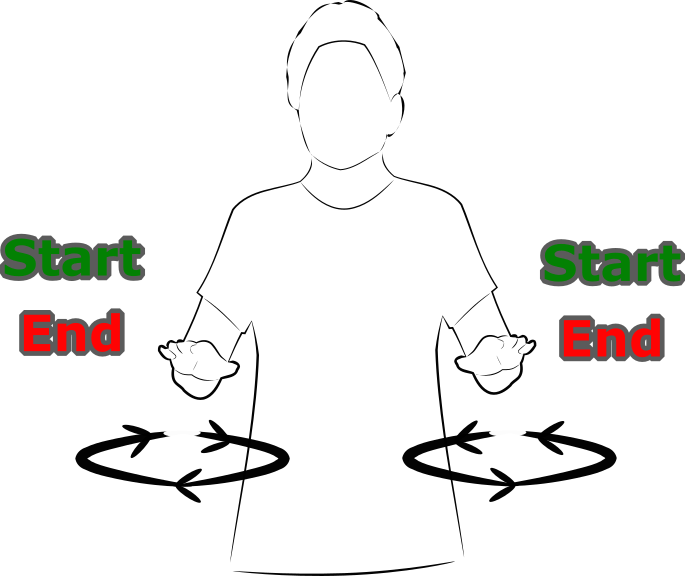}
    \caption{}
    \label{v}
\end{subfigure}
\begin{subfigure}{0.14\textwidth}
    \includegraphics[width=\linewidth]{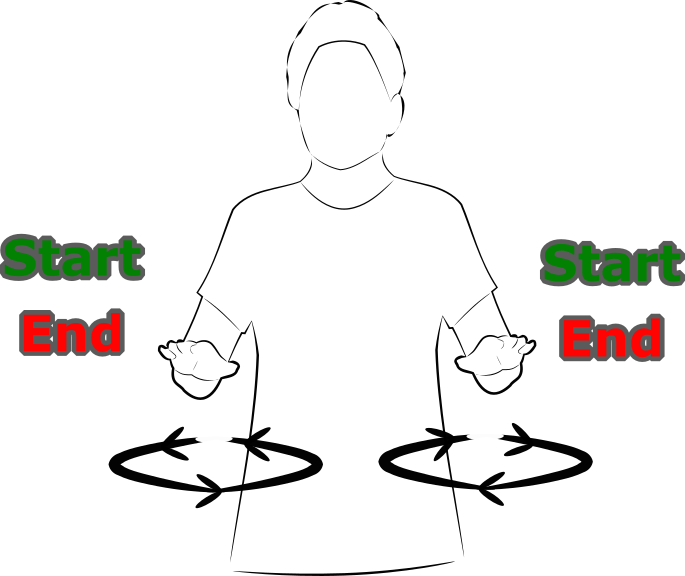}
    \caption{}
    \label{v}
\end{subfigure}
\begin{subfigure}{0.09\textwidth}
    \includegraphics[width=\linewidth]{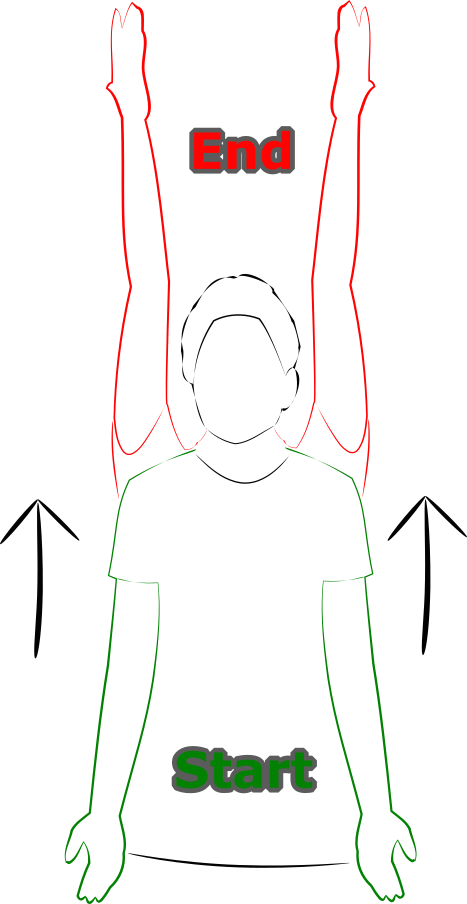}
    \caption{}
    \label{v}
\end{subfigure}
\begin{subfigure}{0.1\textwidth}
    \includegraphics[width=\linewidth]{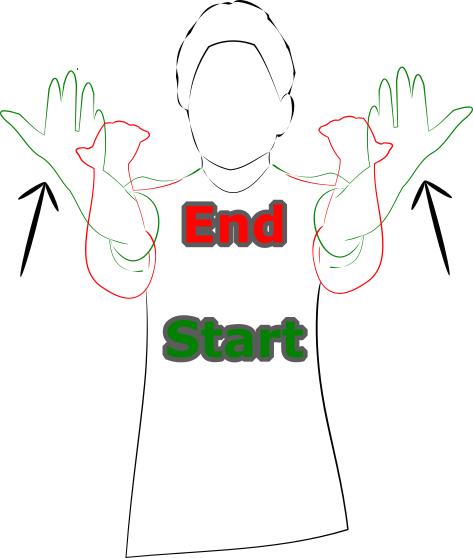}
    \caption{}
    \label{v}
\end{subfigure}
\begin{subfigure}{0.1\textwidth}
    \includegraphics[width=\linewidth]{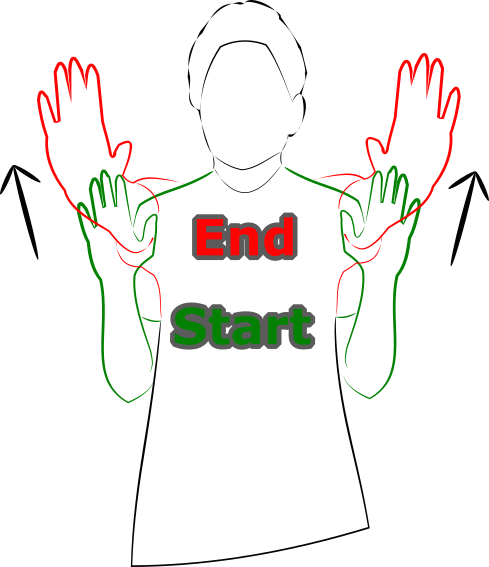}
    \caption{}
    \label{v}
\end{subfigure}
\begin{subfigure}{0.12\textwidth}
    \includegraphics[width=\linewidth]{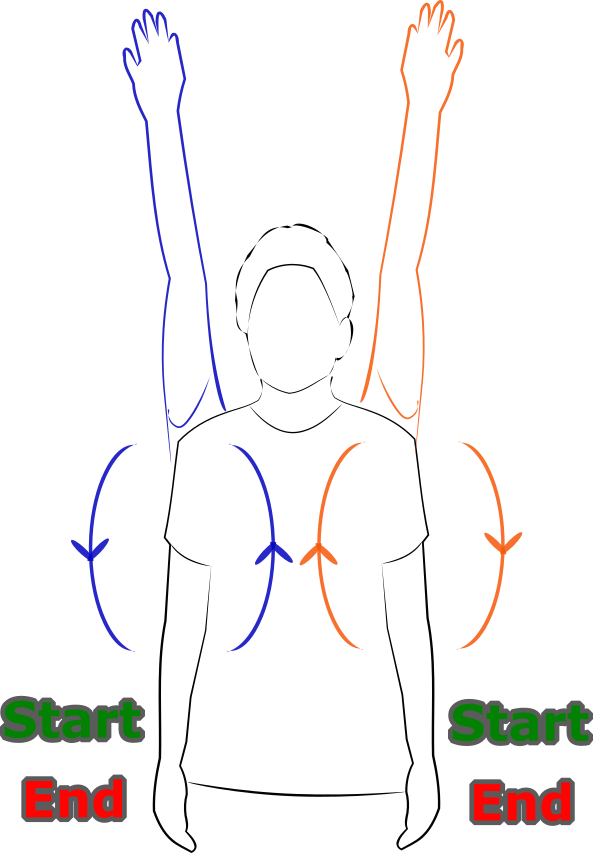}
    \caption{}
    \label{v}
\end{subfigure}
\caption{The gestures performed by participants are (a) 
LD: Lateral Down (1), (b) LF: Lateral to Front (2), (c) LR: Lateral Raise (3), (d) LAC: Left Arm Circle (4), (e) RAC: Right Arm Circle (5), (f) L: Lift (6), (g) Pl: Pull (7), (h) Ps: Push (8), (i) LRo: Left Round (9), (j) RR: Right Round (10), (k) SL: Swipe Left (11), (l) SR: Swipe Right (12), (m) 2HLD: Two Hands Lateral Down (13), (n) 2HLF: Two Hands Lateral to Front (14), (o) 2HLR: Two Hands Lateral Raise (15), (p) 2HIC: Two Hands Inward Circle (16), (q) 2HOC: Two Hands Outward Circle (17), (r) 2HL: Two Hands Lift (18), (s) 2HPl: Two Hands Pull (19), (t) 2HPs: Two Hands Push (20), (u) 2HR: Two Hands Round (21).}
\label{Gestures}
\end{figure*}

The dataset consists of 2,100 samples collected from five participants (2 males, 3 females and heights from 155 cm to 185 cm) standing 3m from the antennas. Each of the 21 unique gestures (which included both simple one-handed and complex two-handed movements, as shown in Fig.~\ref{Gestures}) was repeated 20 times per person.

\subsection{RSS and Phase processing}
The RSS data is normalized via Min-Max normalization:
\begin{equation}\label{eq: RSSNormalization}
[{\mathbf{r}}_n]_m \leftarrow \frac{[{\mathbf{r}}_n]_m - \min({\mathbf{r}}_n)}{\max({\mathbf{r}}_n) - \min({\mathbf{r}}_n)} \quad \forall m= \{1, 2, \dots, M_n\}.
\end{equation}
in which $[{\mathbf{r}}_n]_m$ denotes $m$th RSS reading from $n$th tag and $M_n$ is the number of readings corresponding to the $n$th tag.
Phase unwrapping is applied followed by Savitzky-Golay (S-G)~\cite{schafer2011savitzky}  and Gaussian filtering~\cite{marr1980theory} to remove high-frequency noise.
Both RSS and Phase frames are resampled to an identical length.

\subsection{AoA estimation}
For AoA estimation, we apply the MUltiple SIgnal Classification (MUSIC). After preprocessing the IQ samples and extracting the data corresponding to the target tag (in the case of multiple tags), we obtain the concatenated measurements as given in (\ref{eq:ym-vec}). Since the signal is wide-sense stationary (WSS) and satisfies the second-order ergodicity condition\footnote{
  $\mathbf{\hat{R}}_i \triangleq \mathbb{E}\{\mathbf{Y}_i \mathbf{Y}_i^\mathsf{H}\} \approx \frac{1}{K/4} \mathbf{Y}_i \mathbf{Y}_i^\mathsf{H}$
}, the covariance matrix $\mathbf{\hat{R}_i} \in \mathbb{C}^{2 \times 2}$ of $\mathbf{Y}_i$ is estimated as:

\begin{align}\label{eq:covar}
    \mathbf{\hat R}_i\triangleq \frac{1}{K/4} \ \mathbf{Y}_i \mathbf{Y}_i^{\mathsf{H}}\ \overset{(a)}{=} \alpha \boldsymbol{A}(\theta_\text{i}) \mathbf{R}^i_{\text{ss}} \boldsymbol{A}^{\mathsf{H}}(\theta_\text{i})+\mathbf{R}^i_{\text{n}} , 
\end{align}
where $\mathbf{R}^i_{\text{ss}} \triangleq \frac{1}{K/4} \mathbf{S}\mathbf{S}^{\mathsf{H}}$, 
$\alpha \triangleq g_i^2 P$, and $\mathbf{R}^i_{\text{n}}\triangleq \frac{1}{K/4} \boldsymbol{N}_i\boldsymbol{N}_i^{\mathsf{H}} $. (a) indicates that the noise and the signal are uncorrelated. Next, we perform eigen decomposition on $\mathbf{\hat R}_i$:
\begin{equation}\label{eq:decomposition}
    \mathbf{\hat R}_i\mathbf{u}_r={\lambda}_{r}\mathbf{u}_r, \quad \forall r\in\{n,s\},
\end{equation}
In which $\mathbf{u}_r$ is the eigenvector corresponding to the eigenvalue $\mathbf{\lambda}_r$, with $\mathbf{u}_{\text{n}} <\mathbf{u}_{\text{s}}$. Typically, eigenvectors associated with smaller eigenvalues ($ \mathbf{u}_{\text{n}}$) and larger eigenvalues ($ \mathbf{u}_{\text{s}}$) correspond to the noise subspace and signal subspace, respectively, which are orthogonal to each other. Considering these two orthogonal subspaces, we define the spectrum function, known as the MUSIC estimator, as follows~\cite[Eq. (6)]{schmidt1986multiple}:

\begin{equation}\label{eq:Music}
    P_{_\text{MUSIC}} (\theta_{\text{i}})=\frac{1}{\boldsymbol{a}^{\mathsf{H}}(\theta_{\text{i}}) \mathbf{u}_\text{n}  \mathbf{u}^{\mathsf{H}}_\text{n} \boldsymbol{a}(\theta_{\text{i}})}.
\end{equation}
Finally, we can find the peak of \( P_{\text{MUSIC}} \), which corresponds to the AoA of the \( i \)-th tag, by performing a line search over \( [\theta_{\text{min}}, \theta_{\text{max}}] \), where these values are defined based on our system's field of view. Indeed, our system's field of view is limited because the distance between the antennas is greater than \( \lambda/2 \), which leads to ambiguity in the AoA estimation. \footnote{Despite this limitation, the system setup is configured such that the AoA of the tag, resulting from gesture execution, falls within the unambiguous AoA range.} Estimation is unavailable at certain points due to tag misdetections occurring during gesture execution. Consequently, a Kalman smoother is applied to the AoA estimates produced by the MUSIC algorithm, exploiting the underlying system dynamics.

\subsection{Feature Extraction}\label{FeatureExtraction}
We extract statistical features from signals, i.e., RSS, phase, and AoA, including the mode, the median, the first quartile, the third quartile, the mean, the max, the min, the range, the variance, the standard deviation, the third-order central moment, the kurtosis, the skewness, and the entropy. On top of that, Pearson correlations are extracted between every pair of RSS, phase, and AoA data corresponding to the two tags.
Moreover, we use discrete wavelet transform with Daubechies wavelet and extract low-frequency coefficients.
We apply feature extraction on the processed data in three ways, which are considered to be performed in three separate constrained devices:
\begin{enumerate}
\item Extracting statistical features from both RSS and Phase, and concatenating them (\textbf{SPR})
\item Extracting statistical features from AoA (\textbf{SA})
\item Extracting wavelet coefficients from AoA (\textbf{WA})
\end{enumerate}
Each constrained device uses its extracted features to train an Einsum Network. During testing, the outputs of all constrained devices are combined at the prediction level to determine the final label.

\section{Accuracy Performance}

\subsection{Probabilistic Approach using the Einsum network}
We trained three Einsum Network models independently for each constrained device, using \textbf{SPR}, \textbf{SA}, and \textbf{WA} as input features. These models are denoted as $M_{\text{SPR}}$, $M_{\text{SA}}$, and $M_{\text{WA}}$, respectively. Each Einsum Network outputs a probability distribution over 21 gesture classes, i.e.,
\begin{equation}
M_{f}(g_i) = P(g_i \mid f), \quad i = 1, \ldots, 21,
\end{equation}
where $f \in \{\text{SPR}, \text{SA}, \text{WA}\}$ represents the feature type. 
To obtain the final gesture prediction, we compute the joint probability across the three models and select the gesture with the maximum joint likelihood:
\begin{equation}
\hat{g} = \arg\max_{g_i} \; P(g_i \mid \text{SPR}) \, P(g_i \mid \text{SA}) \, P(g_i \mid \text{WA}).
\end{equation}
This fusion approach leverages the complementary information from multiple feature modalities while preserving the probabilistic interpretability of each Einsum Network.

\begin{figure*}
\centering 
\begin{subfigure}{0.32\textwidth}
    \centering
    \includegraphics[width=\linewidth]{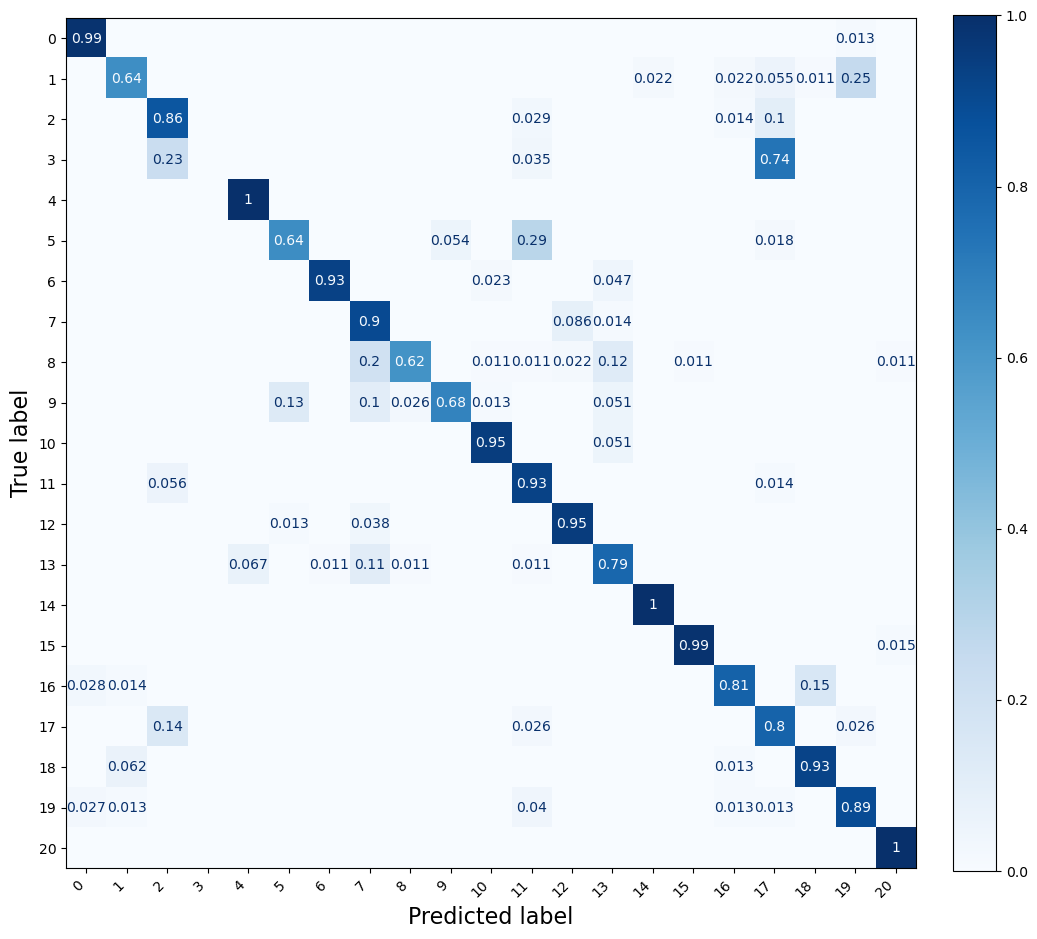}
    \caption{Accuracy=82.8\%}
    \label{StatRSSPhase}
\end{subfigure}
\hfill
\begin{subfigure}{0.32\textwidth}
    \centering
    \includegraphics[width=\linewidth]{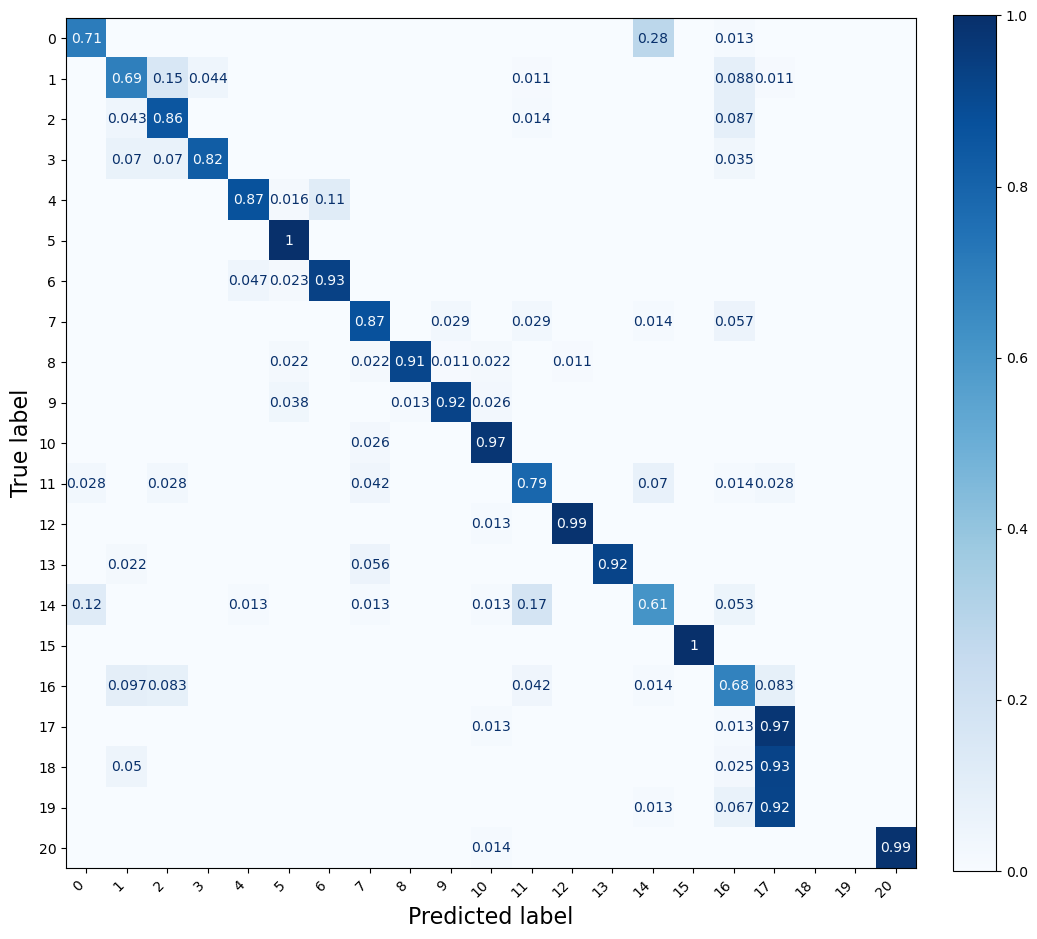}
    \caption{Accuracy=78.21\%}
    \label{StatAoA}
\end{subfigure}
\hfill
\begin{subfigure}{0.32\textwidth}
    \centering
    \includegraphics[width=\linewidth]{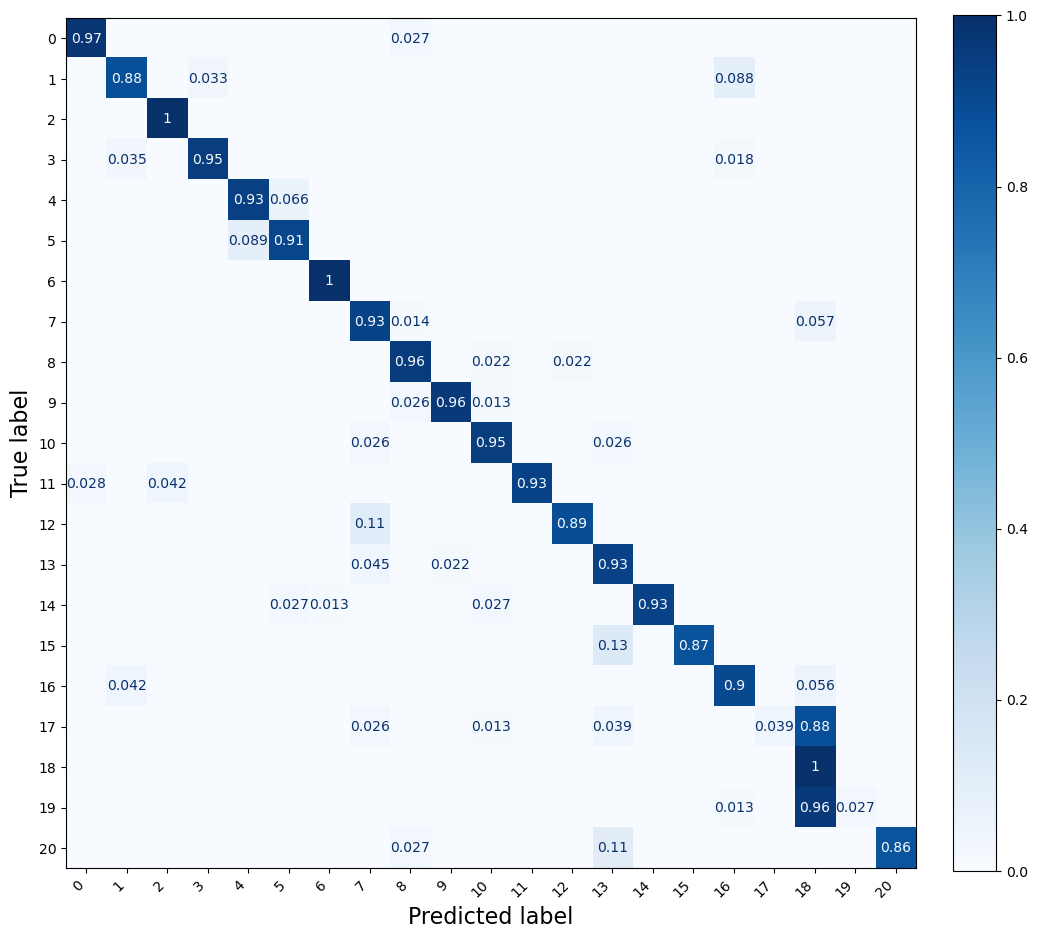}
    \caption{Accuracy=84.84\%}
    \label{WaveAoA}
\end{subfigure}
\caption{Normalized confusion matrices for a single Einsum network using: (a) statistical features from RSS and phase; (b) statistical features extracted from AoA; (c) wavelet-based features derived from AoA.
}
\label{SeparateConfs}
\end{figure*}

\begin{figure}
\centerline{\includegraphics[width=0.8\columnwidth]{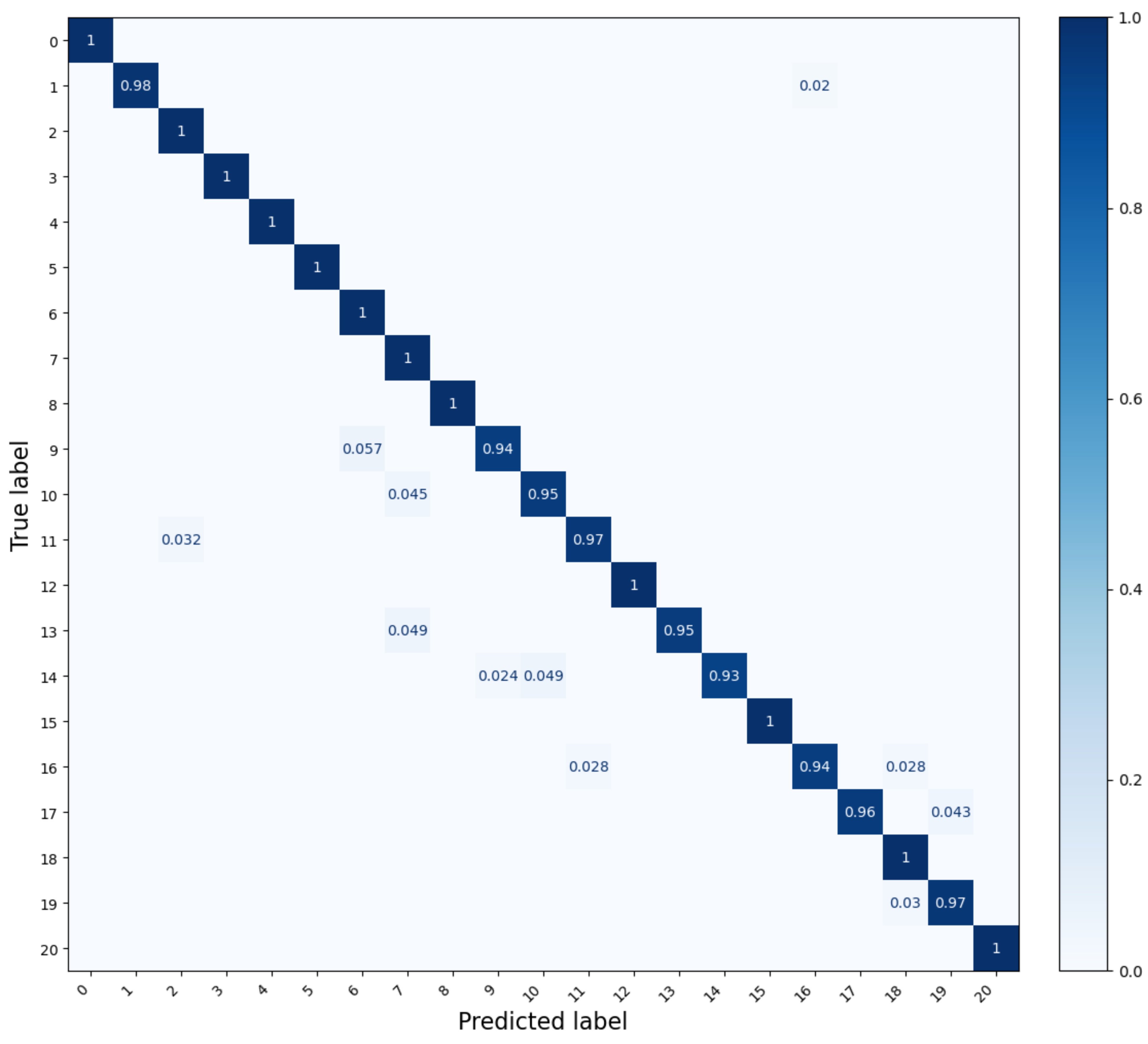}}
\caption{Normalized confusion matrix obtained by averaging the outputs of constrained devices in the decision aggregation module (Merged Einsum Networks), achieving an accuracy of 97.96\%}
\label{Conf}
\end{figure}

Fig~\ref{SeparateConfs} presents the confusion matrices obtained by training individual Einsum networks with different feature extraction strategies. Using wavelet coefficients from the AoA signal yields the highest accuracy of 84.84\%, outperforming the models based on statistical features from RSS and phase (82.8\%) and from AoA (78.21\%). Fig~\ref{Conf} shows the confusion matrix obtained by merging weak learners from individual constrained devices at the prediction level, resulting in an accuracy of 97.96\%. Twelve gestures are perfectly classified, with the others achieving above 93\% accuracy.
\subsection{Comparison with other methods}
The accuracy, precision, recall, and F1-score of our method and the benchmark models are presented in TABLE.~\ref{tabAccuracy}. Deep learning benchmarks were evaluated, including Early Fusion~\cite{calatrava2023light}, which jointly inputs RSS, phase, and AoA signals; Late Fusion~\cite{golipoor2024rfid} and EUIGR~\cite{yu2019rfid}, which integrate features extracted by separate networks. 
Following the approach in~\cite{zhang2022real}, Random Forest Classifiers and Support Vector Machines (SVM)~\cite{cortes1995support} were trained using the separate extracted features detailed in section.~\ref{FeatureExtraction}. The merged Random Forest classifiers and merged SVMs models denote the fusion of their corresponding classifiers at the prediction level. Using the extracted features described in section.~\ref{FeatureExtraction}, three separate Multilayer Perceptron (MLP) networks were trained. The merged MLPs model, as reported in TABLE~\ref{tabAccuracy}, refers to the configuration in which the outputs of these networks are combined at the prediction level. In addition, GRfid~\cite{zou2016grfid}, a dynamic time warping–based AoA method, was applied.

Except for the Merged Random Forest classifiers models, which achieve slightly higher accuracy, our  model (Merged Einsum Networks) outperforms the remaining approaches in terms of accuracy. 



\begin{table}
\caption{Accuracy (Acc.), Precision (Pre.), Recall (Rec.), and F1-score (F1) of methods using phase, RSS, and AoA.}   
\centering
\setlength\tabcolsep{6.5pt} 
\begin{tabular}{|c|c|c|c|c|}
\hline
\multicolumn{1}{|c|}{\diagbox[width=11em]{\textbf{Method}}{\textbf{Metric}}} &  \textbf{Acc.} & \textbf{Pre.} & \textbf{Rec.} & \textbf{F1} \\ \hline
Random Forest with \textbf{SA}~[\citenum{zhang2022real}] & $90.57$ & $91.29$ & $90.57$ & $90.59$ \\ \hline
Random Forest with \textbf{WA}~\cite{zhang2022real} & $92.35$ & $92.71$ & $92.35$ & $92.26$ \\ \hline
Random Forest with \textbf{SPR}~\cite{zhang2022real} & $92.22$ & $92.75$ & $92.21$ & $92.25$ \\ \hline
Merged Random Forest & $98.34$ & $98.49$ & $98.34$ & $98.36$ \\ \hline
SVM with \textbf{SA} & $67.38$ & $67.04$ & $67.38$ & $65.92$ \\ \hline
SVM with \textbf{WA} & $85.09$ & $87.02$ & $85.09$ & $85.1$ \\ \hline
SVM with \textbf{SPR} & $57.32$ & $65.1$ & $57.33$ & $56.2$ \\ \hline
Merged SVMs & $80.25$ & $82.63$ & $80.25$ & $79.33$ \\ \hline
Early Fusion [\citenum{calatrava2023light}]  & $96.94$ & $97.26$ & $96.94$ & $96.95$ \\ \hline
Late Fusion [\citenum{golipoor2024rfid}]  & $95.92$ & $96.21$ & $95.9$ & $95.86$ \\ \hline
EUIGR [\citenum{yu2019rfid}] & $95.41$ & $95.76$ & $95.41$ & $95.39$ \\ \hline
GRfid [\citenum{zou2016grfid}] & $54.42$ & $59.18$ & $54.42$ & $52.15$ \\ \hline
Merged MLPs & $96.24$ & $96.35$ & $96.24$ & $96.23$ \\ \hline
\textbf{Merged Einsum Networks} & $\textbf{97.96}$ & $\textbf{98.05}$ & $\textbf{97.96}$ & $\textbf{97.96}$ \\ \hline
\end{tabular}
\label{tabAccuracy}
\end{table}

\section{Computation Complexity Comparison}
In this section, we compare the computational cost of different approaches. 
The first approach is a deep learning–based model that integrates feature extraction within the network itself. 
The second approach is our merged Einsum Network model method, which operates after an explicit feature extraction stage. 
For comparison, we also evaluate the Random Forest classifier and MLP methods combined with the same feature extraction process, as they demonstrated strong classification accuracy in our experiments.

\subsection{Deep Neural network based methods}
The DNN models used in this work include \textbf{Early Fusion}, \textbf{Late Fusion}, and \textbf{EUIGR}. 
All three are based on a convolutional neural network architecture, in which Convolution (Conv) layers and Fully Connected (FC) layers contribute the majority of computational cost. 
Depending on whether the Conv layer is 1D, 2D, or 3D, the number of Multiplication accumulation operations (MACs) can be computed as follows:

\begin{equation}
\text{MAC}_{\text{Conv}} =  D_o \cdot H_o \cdot W_o \cdot C_o \cdot C_i \cdot K_D \cdot K_H \cdot K_W,
\end{equation}
where $K_D, K_H, K_W$ denote the kernel sizes, $C_i$ and $C_o$ are the numbers of input and output channels, and $D_o, H_o, W_o$ represent the spatial dimensions of the output feature map (for 1D or 2D convolutions, the unused dimensions are omitted).

The number of MACs in a FC layer is given by:
\begin{equation}
\text{MAC}_{\text{FC}} = N_{\text{i}} \cdot N_{\text{o}},
\end{equation}
where $N_i$ and $N_o$ are the numbers of input and output neurons, respectively.

Table \ref{dnnmac} details the parameters of each Conv and FC layer of 3 models and the total MAC operations.

\begin{table}
\centering
\caption{Summary of MAC operations for Early Fusion, Late Fusion, and EUIGR architectures. Conv3D parameter contains $(D_o,H_o,W_o,C_o,C_i,K_D,K_H,K_W)$; Conv2D parameter contains $(H_o,W_o,C_o,C_i,K_H,K_W)$; FC parameter contains $N_i, N_o$.}
\label{tab:MAC_summary}
\renewcommand{\arraystretch}{1.15}
\setlength{\tabcolsep}{5pt}
\begin{tabular}{l l c c}
\hline
\textbf{Architecture} & \textbf{Layer Type} & \textbf{Parameters} & \textbf{Count} \\
\hline
\multirow{3}{*}{\textbf{Early Fusion}} 
& Conv3D & $ (2,5,35,64,1,5,5,5)$ & 1 \\
& Conv3D & $(2,5,35,64,64,5,5,5)$ & 8 \\
& FC & $(64,21)$ & 1 \\
\hline
\multirow{7}{*}{\textbf{Late Fusion}} 
& Conv2D & $(2,2,64,35,5,5)$ & 2 \\
& Conv2D & $(2,1,64,35,5,5)$ & 1 \\
& Conv2D & $(2,2,64,64,5,5)$ & 2 \\
& Conv2D & $(2,1,64,64,5,5)$ & 1 \\
& Conv2D & $(1,1,64,64,5,5)$ & 3 \\
& FC & $(192,64)$ & 1 \\
& FC & $(64,21)$ & 1 \\
\hline
\multirow{9}{*}{\textbf{EUIGR}} 
& Conv2D & $(2,2,64,35,5,5)$ & 4 \\
& Conv2D & $(2,1,64,35,5,5)$ & 2 \\
& Conv2D & $(2,2,64,64,5,5)$ & 4 \\
& Conv2D & $(2,1,64,64,5,5)$ & 2 \\
& Conv2D & $(1,1,64,64,5,5)$ & 6 \\
& FC & $(192,64)$ & 2 \\
& FC & $(64,21)$ & 1 \\
& FC & $(64,9)$ & 1 \\
\hline
\multicolumn{4}{l}{\textbf{Total MACs per Architecture:}} \\
\multicolumn{2}{l}{Early Fusion} & \multicolumn{2}{l}{\textbf{1.4 $\times$ 10\textsuperscript{9}}} \\
\multicolumn{2}{l}{Late Fusion} & \multicolumn{2}{l}{\textbf{1.9 $\times$ 10\textsuperscript{6}}} \\
\multicolumn{2}{l}{EUIGR} & \multicolumn{2}{l}{\textbf{3.8 $\times$ 10\textsuperscript{6}}} \\
\hline
\end{tabular}
\label{dnnmac}
\end{table}

\subsection{Probabilistic Approach using Einsum network}
In our  Einsum Network-based approach, the computation cost is shared by the feature extraction and inference costs. 

\subsubsection{Feature Extraction Costs}\label{ExtractionCosts}
As described in Section~\ref{FeatureExtraction}, the \textbf{SPR} module extracts 14 statistical features from both RSS and Phase signals (35 samples each) and concatenates results from eight signals (two tags and two antennas). 
Four Pearson correlations are additionally computed between corresponding RSS and Phase pairs, yielding a total of $116$ features. 
The estimated cost is $8\times35\times14=3920$ operations (OPs) for feature extraction and $4\times35=140$ OPs for correlations.

The \textbf{SA} module extracts 14 statistical features from two AoA signals (35 samples each) and computes one correlation between $(\text{AoA1}, \text{AoA2})$, resulting in $29$ features. 
Its cost is $2\times35\times14=980$ OPs for feature extraction and $35$ OPs for correlation.

The \textbf{WA} module extracts wavelet-based features from two AoA signals (35 samples each) using a Daubechies-2 ($\text{db2}$) wavelet with single-level decomposition. 
Only approximation coefficients are retained, giving $19\times2=38$ features. 
The $\text{db2}$ wavelet has a filter length of four, resulting in approximately $2\times35\times4=280$ OPs.

\begin{table}
\caption{Feature extraction cost per module.}
\centering
\setlength\tabcolsep{4pt}
\begin{tabular}{lccc}
\hline
\textbf{Module} & \textbf{Features} & \textbf{Correlations} & \textbf{Total OPs} \\ \hline
SPR & 116 & 4 & $4060$ \\
SA  & 29  & 1 & $1015$ \\
WA  & 38  & -- & $280$ \\ \hline
\textbf{Total} & -- & -- & \textbf{5355} \\ \hline
\end{tabular}
\label{tab:feature_costs}
\end{table}

Overall, the total estimated computational cost for feature extraction across all modules is approximately $5355$ OPs, as shown in Table~\ref{tab:feature_costs}.

\subsubsection{Einsum Network Model Costs}
The Einsum Network model is characterized by several structural parameters: depth $D$, number of sum constrained devices in each Einsum $K$, number of leaf distributions $L$, number of repetitions $R$, and number of classes $C$. 

Each Einsum operation with $K$ sum constrained devices performs $S_k = W_{kij}U_iV_j$, where $W$ is a $K\times K\times K$ tensor and $U,V$ are $K$-dimensional vectors. This requires $2K$ multiplications and $K^3$ MACs, approximated as $(K^3+K)$ MACs per Einsum. Each model starts with a single Einsum at the root and follows a binary splitting structure, where each layer doubles the number of Einsums relative to the previous one, resulting in $2^d$ Einsums at depth $d$. Summing over all $D$ layers and repeating the structure $R$ times gives the total Einsum cost:
\begin{equation}
\text{MAC}_{\text{einsum}} 
= \left(\sum_{d=1}^{D} 2^d\right) \cdot R \cdot (K^3 + K)
= (2^{D+1} - 2) \cdot R \cdot (K^3 + K).
\label{eq:einsum}
\end{equation}

Each leaf performs normalization $z=(x-\mu)\sigma^{-1}$, involving one subtraction and one multiplication per input, thus the total leaf operations are
\begin{equation}
\text{MAC}_{\text{leaf}} = 2^D.
\label{eq:leaf}
\end{equation}

The root mixing operation is analogous to a fully connected layer, where $R$ and $C$ correspond to input and output dimensions:
\begin{equation}
\text{MAC}_{\text{mix}} = R \times C.
\label{eq:mix}
\end{equation}

Each model and the total costs are detailed in Table~\ref{tab:Einsum Network_cost}. 
We can see that the $\text{MAC}_{\text{einsum}}$ term dominates the computation.

\begin{table}
\caption{Computation breakdown of Einsum Network model configurations and corresponding MAC operations (analytical estimate).}
\centering
\setlength\tabcolsep{5pt}
\begin{tabular}{lccccc}
\hline
\textbf{Model} & \textbf{$(D,K,L,R,C)$} & $\text{MAC}_{\text{einsum}}$ & $\text{MAC}_{\text{leaf}}$ & $\text{MAC}_{\text{mix}}$ & \textbf{MACs} \\ \hline
SPR & (6, 2, 10, 10, 21) & $12600$ & $128$ & $210$ & $12938$ \\
SA  & (4, 2, 10, 10, 21) & $3000$  & $32$  & $210$ & $3242$  \\
WA  & (5, 2, 10, 10, 21) & $6200$  & $64$  & $210$ & $6474$  \\ \hline
\textbf{Total} & -- & \textbf{21800} & \textbf{224} & \textbf{630} & \textbf{22654} \\ \hline
\end{tabular}
\label{tab:Einsum Network_cost}
\end{table}

\subsection{Merge-based Approach using Random Forest classifier and MLP}
We analyze the computational cost of the \textbf{Random Forest classifier} and the \textbf{MLP}. The feature extraction cost remains identical to the Einsum Network-based pipeline and is therefore omitted here.

\textbf{Random Forest classifier:} The inference cost of Random Forest classifier is estimated in MAC-equivalent operations, where each arithmetic or comparison step counts as one operation. 
Each internal node performs a feature read and threshold comparison ($x_f<t$), requiring two operations per level. 
Leaf constrained devices compute class probabilities from stored counts, followed by aggregation, averaging, and final $\mathrm{argmax}$ classification. 
With $n_{\text{estimators}}{=}100$, $\text{max\_depth}{=}10$, and $C{=}21$, the total per-forest cost is approximately $8{,}241$ MAC-equivalent operations, as summarized in Table~\ref{tab:Random Forest classifier_cost}, yielding $3\times8{,}241=24{,}723$ operations across three Random Forest classifier models.

\begin{table}
\caption{Estimated Random Forest classifier inference cost.}
\centering
\setlength\tabcolsep{6pt}
\begin{tabular}{lcc}
\hline
\textbf{Component} & \textbf{Formula} & \textbf{Ops (MAC-equiv.)} \\ \hline
Node traversal & $2\times100\times10$ & $2000$ \\
Leaf normalization & $100\times(2C{-}1)$ & $4100$ \\
Aggregation \& averaging & $100\times C + C$ & $2121$ \\
Argmax & $(C{-}1)$ & $20$ \\ \hline
\textbf{Per forest total} & -- & \textbf{8241} \\
\textbf{All 3 models} & $3\times8241$ & \textbf{24723} \\ \hline
\end{tabular}
\label{tab:Random Forest classifier_cost}
\end{table}

\textbf{MLP:} The inference cost of an MLP is dominated by dense FC layer multiplications, computed as
\begin{equation}
\text{MAC}_{\text{MLP}} = \sum_{l=1}^{L-1} N_l N_{l+1},
\end{equation}
where $L$ is the number of layers, and $N_l$ and $N_{l+1}$ denote the number of neurons in consecutive layers. The final per-model MAC counts are summarized in Table~\ref{tab:mlp_macs}.

\begin{table}
\caption{Total MACs for the MLP-based models.}
\centering
\setlength\tabcolsep{8pt}
\begin{tabular}{lcc}
\hline
\textbf{Model} & \textbf{Parameters $(N_1,\dots,N_L)$} & \textbf{Total MACs} \\ \hline
SPR & $(116,\,128,\,21)$ & $17{,}536$ \\ 
SA & $(29,\,32,\,21)$ & $1{,}600$ \\
WA & $(38,\,128,\,21)$ & $7{,}552$ \\
\hline
\textbf{Total} & - &  \textbf{26688}\\ \hline
\end{tabular}
\label{tab:mlp_macs}
\end{table}

\section{Discussion of Accuracy and Cost}
We compare three DNN architectures and three merge-based models in terms of classification accuracy and computational cost. 
Accuracy corresponds to the final test performance, while cost is expressed in total MAC-equivalent operations per sample, combining both feature extraction and model inference.

As shown in Table~\ref{tab:accuracy_cost}, the DNN architectures achieve strong accuracies but incur significantly higher computational costs, with the Early Fusion model requiring approximately $1.4\times10^9$ MACs. 
In contrast, the  Merge-based approaches achieve comparable or superior accuracies with drastically reduced complexity. 

Among the Merge-based models, \textbf{Merged Einsum Networks} achieves $97.96\%$ accuracy with only $22{,}654 + 5{,}355$ MACs, providing an excellent trade-off between efficiency and interpretability. 
As a probabilistic and explainable approach, it allows transparent reasoning about feature contributions and uncertainty estimation capabilities typically absent in conventional neural networks. 
\textbf{Merged MLPs} attains accuracy $96.24\%$ with slightly higher cost but lacks interpretability. 
\textbf{Merged Random Forest classifiers} achieves the highest accuracy ($98.34\%$) but presents hardware limitations due to its irregular control flow and memory-intensive tree traversal, making it less suitable for efficient hardware acceleration compared to Einsum Network or MLP. 

Using the efficiency metric defined as~\cite{cueto2024framework}: 
\begin{equation}\label{Efficiency}
    \text{efficiency}_\text{Score}=\frac{\text{Accuracy}}{log (\text{MACs})},
\end{equation}
the \textbf{Merged Einsum Networks} and \textbf{Merged Random Forest classifiers} achieve scores of 22.5\% and 22.4\%, respectively, indicating that both models are highly efficient. If hardware-efficient acceleration is required, the \textbf{Merged Einsum Networks} model is the preferred choice. Conversely, in the absence of hardware constraints, the \textbf{Merged Random Forest classifiers} provides slightly better performance.

Overall, the Merge-based models provide substantial computational savings while maintaining high accuracy. 
Among them, \textbf{Merged Einsum Networks} stands out as the most balanced and trustworthy solution, combining interpretability, low computational cost, and suitability for real-time, low-power applications.

\begin{table}
\caption{Comparison of model accuracy and computational cost. All 3 merged models have an extra feature extraction cost, see \ref{ExtractionCosts}.}
\centering
\setlength\tabcolsep{6pt}
\begin{tabular}{lcc}
\hline
\textbf{Model} & \textbf{Accuracy (\%)} & \textbf{Total Cost (MACs)} \\ \hline
Early Fusion DNN & 96.94 & $1.4\times10^9$ \\
Late Fusion DNN  & 95.92 & $1.9\times10^6$ \\
EUIGR DNN        & 95.41 & $3.8\times10^6$ \\ 
Merged Einsum Networks     & 97.96 & $22{,}654 $ \\
Merged MLPs       & 96.24 & $26{,}688 $ \\
Merged Random Forest classifiers       & 98.34 & $24{,}723 $ \\ \hline
\end{tabular}
\label{tab:accuracy_cost}
\end{table}

\section{Conclusion}
We have proposed a gesture-recognition pipeline for networks of distributed, resource constrained devices utilising Einsum Networks. 
The system has been validated for a specialized accelerator hardware in a scenario of low-power, body-worn passive backscatter-type tags for gesture recognition. 
The  model employs dedicated, energy-efficient probabilistic circuit hardware for feature processing (RSS, phase, AoA) before fusing their outputs for gesture prediction. 
Experimental validation confirmed that this method outperforms most benchmark models in recognition accuracy, achieving 97.96\%, and also achieves substantially lower computational complexity.
Deploying a single Einsum network fed by extracted statistical features or wavelet coefficients yielded a maximum accuracy of 84.84\%. 
These findings show the potential of integrating such architecture into next-generation embedded systems, where sensors act as mini readers capable of locally collecting, processing, and performing the classification task.

\appendices


\ifCLASSOPTIONcompsoc
\else


\ifCLASSOPTIONcaptionsoff
  \newpage
\fi

\balance 
\bibliographystyle{ieeetr}
\bibliography{ref}

\end{document}